# An Integral Experiment on Polyethylene Using Radiative Capture in Indium Foils in a High Flux D-D Neutron Generator.

Nnaemeka Nnamani[a]*, Karl van Bibber[b,a], Lee A Bernstein[b], Jasmina L Vujic[a], Jonathan T. Morrell[a]

[a]University of California Berkeley, Nuclear Engineering, Berkeley, California 94720.

[b]Lawrence Berkeley National Laboratory, Berkeley, California 94720

*E-mail: nnamani.nnaemeka@berkeley.edu

# An Integral Experiment on Polyethylene Using Radiative Capture in Indium Foils in a High Flux D-D Neutron Generator.

The Department of Nuclear Engineering, University of California Berkeley built a D-D neutron generator called the High Flux Neutron Generator (HFNG). It operates in the range of 100-125 keV of accelerating voltage. The generator produces neutron current of about $10^8$ per second. These neutrons have energies between 2.2 – 2.8 MeV.

We report here the results of a measurement of the scattered vs unscattered neutron fluence on polyethylene determined via neutron activation of multiple natural indium foils from a D-D neutron generator. Both the angle-integrated spectrum and the angle differential results are consistent with the predictions of the Monte Carlo N-Particle Transport (MCNP) code, using the ENDF/B-VII.1. This supports shielding calculations in the fast energy region with high density polyethylene (HDPE). To the best of our knowledge no integral benchmark experiment has been performed on polyethylene using $D(D,n)^3He$ neutron spectrum.

Keywords: Cross-sections; nuclear data; integral experiments; neutron, polyethylene.

I. **MOTIVATION**

Cross-sections nuclear data are invaluable to the nuclear community. In addition to providing information on nuclear structure of matter and reaction dynamics, they provide practical support for applications including design and modelling of shielding geometry for radiation sources such as radiation treatment machines, etc.

Integral benchmark experiments are used to test the accuracy and reliability of evaluated nuclear data over a range of particle energies and angles. They are a rich source of information with which a wide range of validation and comparison exercises can be made [1]. Such benchmark experiments provide global measures of data performance for applications and are a valuable resource for nuclear data testing and evaluation efforts [2]. Inconsistencies in the basic nuclear data file can then be determined and feedback given to the evaluator, thus providing an iterative approach to correct, modify, and improve the recommended nuclear data in the files [3], making them an essential part of the nuclear data evaluation process. As a result, integral experiments play an essential role in nuclear data validation and improvements[4].

Integral benchmark data sets are particularly valuable for materials that are commonly used in the fabrication and shielding of nuclear energy systems. Polyethylene is an important material in shielding of humans from the neutron radiation due to the moderating ability of hydrogen. The polyethylene is often used in conjunction with materials that have a high thermal neutron capture cross section such as boron, cadmium and gadolinium, in order to absorb the moderated neutrons. In most cases these high thermal neutron cross section materials are dispersed inside polyethylene. To determine if the evaluated nuclear data for the constituent elements in polyethylene accurately reproduce what is observed in a shielding experiment, Monte Carlo simulations of radiative capture and inelastic scattering reaction rates carried out using MCNP [5] were compared to values measured using the High Flux Neutron Generator (HFNG) on the UC- Berkeley campus[6]. To the best of our knowledge no integral benchmark experiment has been performed on polyethylene using $D(D,n)^3He$ neutron spectrum.

## II. THE EXPERIMENT

The experiment was done at a High Flux Neutron Generator (HFNG) in the Department of Nuclear Engineering, University of California at Berkeley. The HFNG produces neutrons via D(D,n)alpha reaction. The geometry of the HFNG is shown in Figure *1* and described in [7]. The HFNG consists of ion sources, a self-loading titanium target and aluminium chamber. The target is enclosed in the aluminium chamber. Deuterons from the ion source are accelerated onto the self-loading target. Deuterons adsorbed on the target migrate to the surface of the target and interact with incoming deuterons to produce neutrons via the D(D,n)alpha reaction. The HFNG was designed to use two ion sources. However, only one of the ion sources was used for this experiment.

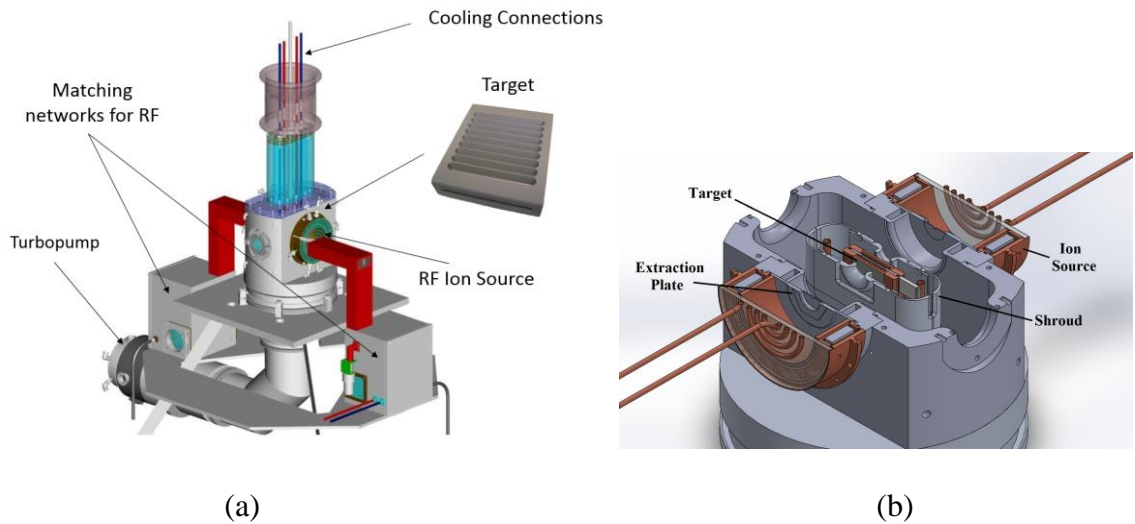

(a)  (b)

Figure 1. An external schematic of the HFNG (a) and a CAD drawing showing many of key features (b) including the neutron production target. The distance between the two ion sources is approximately 14 cm. [7]

Eight natural indium foils of approximate radius 0.45 cm, approximate thickness 0.05 cm and whose masses ranged from 0.0734 - 0.3074 g were put in the target holder made of aluminium with a polyethylene handle as shown in *Figure 2* (b).

Activation of natural indium foils yields the product of interest: $^{116m1}_{49}In$ via radiative capture. The decay of $^{116m1}_{49}In$ (half-life of 54.27 minutes) produces a 416.9 keV gamma ray 27.2% of the time.

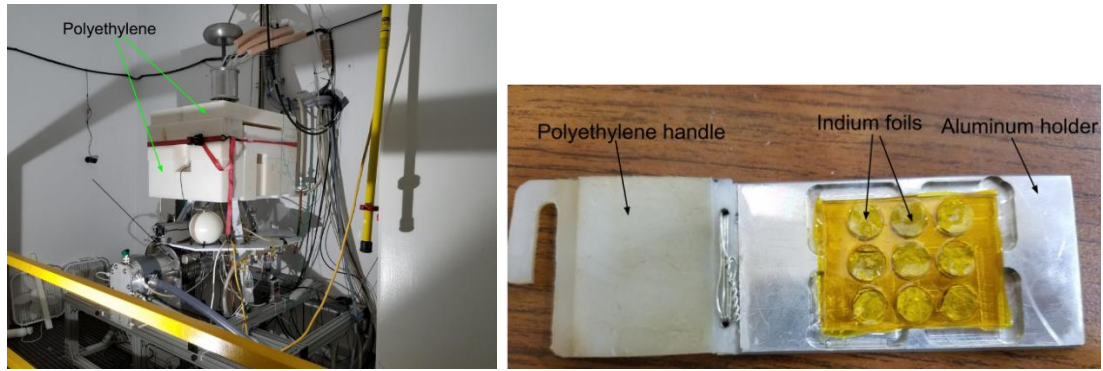

(a) (b)

Figure 2. (a) The HFNG showing the polyethylene cover just before the first irradiation. (b). The 8 indium foils in their aluminum holder. The foils are numbered from left to right.

Two set of experiments were done. In the first experiment the target holder was inserted into the slot at the centre of the titanium target and positioned in such a way that the axis of the central indium foil is along the direction of the deuteron beam. At this position, the foils are approximately 1 cm behind the neutron production area. Plates of polyethylene 20 cm thick were used to cover the chamber as shown in Figure 2 (a). The foils were then irradiated for four hours. At the end of irradiation, the foils were left to cool down for 30 minutes so that activation products with short half-lives would decay away, making it safe to handle the foils.

Gamma-rays from the reaction product were observed using a High Purity Germanium (HPGe) detector. Each foil was counted for 5 minutes because of the short half-life of $^{116m1}_{49}In$.

While counting was being done, a second set of indium foils was irradiated, with the polyethylene plates surrounding the chamber removed. The details of the foils and irradiations are shown in Table 1, Table 2, and Table 3. The uncertainty in waiting time is no more than 60 seconds.

| Experiment | irradiation time (minutes) |
|---|---|
| Without polyethylene plates | 240±1 |
| With polyethylene plates | 240±1 |

Table 1. Irradiation details.

| Foil number | mass(g) | waiting time(s) | counting time(s) |
|---|---|---|---|
| 1 | 0.1477 | 1920 | 400 |
| 2 | 0.1725 | 2460 | 300 |
| 3 | 0.1088 | 3000 | 300 |
| 4 | 0.2131 | 3540 | 300 |
| 5 | 0.0734 | 4080 | 300 |
| 6 | 0.1682 | 4560 | 300 |
| 7 | 0.3074 | 5040 | 300 |
| 8 | 0.1105 | 5520 | 300 |
| 9 | 0.2293 | 5760 | 300 |

Table 2. Foil masses and counting details for the irradiation in which the polyethylene plates are not in place.

| Foil number | mass(g) | Waiting time(s) | counting time(s) |
|---|---|---|---|
| 1 | 0.1227 | 2040 | 300 |
| 2 | 0.1544 | 2460 | 300 |
| 3 | 0.1403 | 2880 | 300 |
| 4 | 0.1161 | 3240 | 300 |
| 5 | 0.1752 | 3660 | 300 |
| 6 | 0.156 | 4080 | 300 |
| 7 | 0.1407 | 4500 | 300 |
| 8 | 0.121 | 4860 | 300 |
| 9 | 0.1136 | 5280 | 300 |

Table 3. Foil masses and counting details for the irradiation in which polyethylene plates are in place.

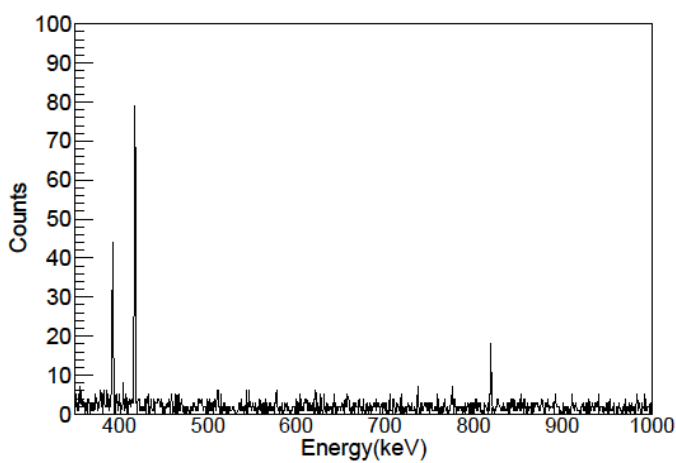

(a)

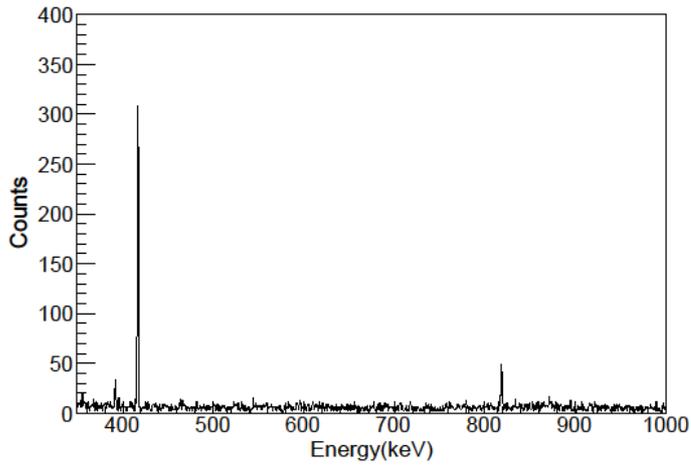

(b)

Figure 3. Gamma-ray from central indium foil showing the 416.9 keV energy peak. (a) no polyethylene plate cover. (b) polyethylene plate cover in position.

The ROOT data analysis code was used to remove the background and then fit a gaussian function to the 416.9 keV energy peak. Fit parameters were then used to calculate the counts. The results of this fit are shown in Figure *4*. These intensities are presented in Table 4, together with the associated statistical error.

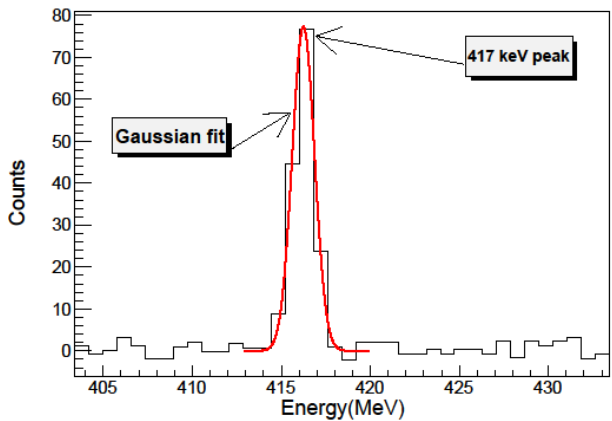

Figure 4. Fitting the 416.9 keV peak with a Gaussian function to find the area.

|             | No poly |       | Poly present |       |
|-------------|---------|-------|--------------|-------|
| Foil number | Counts  | Error | Counts       | Error |
| 1           | 196     | 16    | 460          | 28    |
| 2           | 278     | 23    | 596          | 28    |
| 3           | 117     | 16    | 490          | 28    |
| 4           | 185     | 19    | 386          | 30    |
| 5           | 129     | 16    | 598          | 35    |
| 6           | 188     | 21    | 421          | 28    |
| 7           | 167     | 17    | 306          | 24    |
| 8           | 84      | 10    | 278          | 23    |
| 9           | 136     | 19    | 214          | 19    |

Table 4. Counts of the 416.9 keV gamma peak.

Because the axis of the central indium foil may not coincide with the deuteron beam, a quadratic surface fit was performed to determine the axis of the beam through the foil by making use of quadratic dependence of neutron yield with angle [8]. The result was used in the MCNP calculation to ensure that the simulation is done under the same geometric configuration of the experiment.

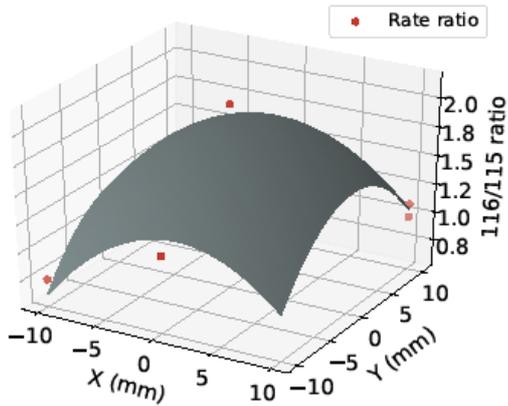

Figure 5. Quadratic surface fit to count data. From this fit, it is determined that the central indium foil is displaced from the origin approximately by x, y value of 1 mm and 0.5 mm respectively.

### III. MCNP simulation

The angular distribution of energy and yield of neutrons in D(D,n)$^3$He reaction as a function of deuteron energy are well known and are described in [8,9]. Using those references, an MCNP input file was built for the HFNG. This is described in [7,10].

A two-dimensional MCNP rendering of the geometry of the D-D neutron generator is shown in Figure *6*.

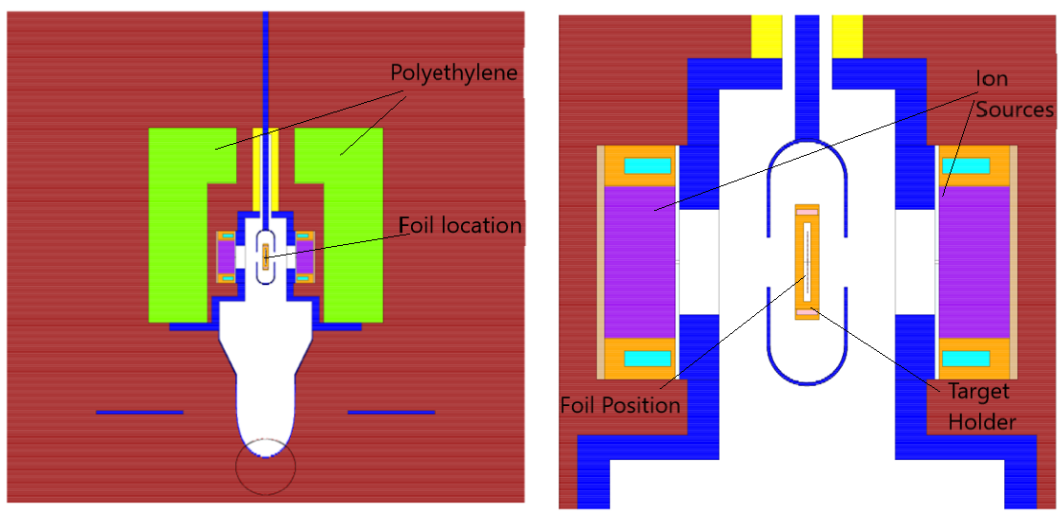

*(a)* *(b)*

Figure 6. MCNP geometry. (a) Side view of the HFNG showing the polyethylene cover (green). (b)Side view of the HFNG showing the ion sources, target and indium foil holder.

The simulation involves calculating the rates of radiative capture to the $^{116m1}$In metastable state with polyethylene plates cover in place as shown in Figure *6* (a), and then without the polyethylene plates cover in place.

Primary neutrons produced in the HFNG undergo radiative capture into the 127.3 keV energy level in $^{116}$In, producing $^{116m1}$In that has a half-life of 54.29 minutes, [11].

$$^{115}_{49}\text{In} + n \rightarrow {}^{116m1}_{49}\text{In} + \gamma \qquad (1)$$

$^{116m1}$In decays by beta emission to $^{116}$Sn with branching ratio of 100%. The $^{116}$Sn is left in an excited state and emits a gamma ray cascade. The last gamma ray in the cascade is the 416.9 keV line with intensity of 27% [12].

The cross section for production of $^{116m1}_{49}$In is 0.79 times the cross section for neutron capture in $^{115}_{49}$In as given in ENDF/B-VII.1 nuclear data [13] retrieved from [14] as given by [15].

A billion neutrons were simulated on the Berkeley Research Computing cluster using 24 core processors. The ratio of rates of formation of $^{116m1}_{49}$In was then extracted from the output and is shown in Table 5.

| Foil number | polyethylene/no polyethylene | Error |
|---|---|---|
| 1 | 5.932 | 0.269 |
| 2 | 5.161 | 0.236 |
| 3 | 6.404 | 0.345 |
| 4 | 4.673 | 0.188 |

| 5 | 3.420 | 0.102 |
|---|---|---|
| 6 | 4.789 | 0.216 |
| 7 | 5.800 | 0.257 |
| 8 | 4.800 | 0.236 |
| 9 | 6.489 | 0.395 |

Table 5. Rate ratios, representing rate of formation of $^{116m1}$In the presence and absence of polyethylene plate covers, at different indium foil position.

## IV. Data analysis

### IV.A. Production and decay of radioactivity

Radioactive isotopes decay at the same time as they are being produced in nuclear reactions. The amount of the radioactive isotope at any time $t$ depends on its production rate $R$ and rate of disintegration A. Let j be the radionuclide:

$$dN_j = R_j dt - A dt \quad (2)$$

(Eq. 2) is the rate equation. It could be re-written using the fact that $A = N_j \lambda_j$, as:

$$\frac{dN_j}{dt} = R_j - \lambda_j \quad (3)$$

This is a first order linear differential equation whose solution yields the concentration of atom $N_j$ at any given time as:

$$N_j = \frac{R_j}{\lambda_j}(1 - \exp(-\lambda_j t))$$

The number of atoms of isotope j at the end of irradiation time $t = t_i$ is:

$$N_{j1} = \frac{R_j}{\lambda_j}(1 - \exp(-\lambda_j t_i)) \quad (4)$$

After irradiation, $N_{j1}$ decays exponentially. Thus, for a waiting time $t_d$, the number of remaining atoms, $N_{j2}$, is given by:

$$N_{j2} = N_{j1} \times \exp(-\lambda_j t_d) = \frac{R_j}{\lambda_j}(1 - \exp(-\lambda_j t_i)) \times \exp(-\lambda_j t_d)$$

If the number of either 416.9 keV gamma rays emitted are counted for a time $t_c$, the number observed is given by:

$$N_{j3} = N_{j2}(1 - \exp(-\lambda_j t_c))$$

$$= \frac{R_j}{\lambda_j}(1 - \exp(-\lambda_j t_i)) \times \exp(-\lambda_j t_d) \times (1 - \exp(-\lambda_j t_c))$$

Since not every decay result in emission of the gamma ray of interest and not every gamma ray emitted by the atoms is detected by the detector, the number of counts is therefore given as:

$$N_\gamma = N_{j3} \times BR \times \epsilon$$

Where BR is the branching ratio of the gamma ray and $\epsilon$ is the absolute full-energy peak efficiency of the detector for the gamma ray.

$$\therefore N_\gamma = \frac{R_j}{\lambda_j}(1 - \exp(-\lambda_j t_i)) \times \exp(-\lambda_j t_d) \times (1 - \exp(-\lambda_j t_c)) \times BR \times \epsilon \quad (5)$$

### IV.B. Validation of Polyethylene Nuclear Data by $^{116m1}$In double ratio method.

Validation of the nuclear data for carbon and hydrogen in polyethylene is done using a double ratio method. First, the ratio of $^{116m1}$In formation rate in the presence and absence of polyethylene plates cover is calculated for both experiment and simulation. The ratio of these two ratios is formed to see how experiment compares to the simulation.

This approach has two advantages. One, the calculation of efficiency is not required since the efficiency term cancels out. Two, branching ratios also cancel out. Therefore, uncertainties from efficiency and branching ratio vanish. Furthermore, the difference in irradiation time is negligible and within the uncertainty in irradiation time. However, since the masses of the indium foils are not the same, the experimental ratio must be normalized by the masses of the indium foils.

Using (Eq. 5), the number of the 416.9 keV gamma rays detected after counting for a time $t_c$ is:

$$N_{416.9\text{ keV}} = \frac{R_{116m1}}{\lambda_{116m1}} (1 - \exp(-\lambda_{116m1} t_i)) \times \exp(-\lambda_{116m1} t_d) \times$$
$$(1 - \exp(-\lambda_{116m1} t_c)) \times BR_{416.9\text{ keV}} \times \epsilon_{416.9\text{ keV}} \qquad (6)$$

And thus

$$R_{116m1} = \frac{N_{416.9\text{ keV}} \times \lambda_{116m1}}{(1 - \exp(-\lambda_{116m1} t_i)) \times \exp(-\lambda_{116m1} t_d)} \times$$
$$\frac{1}{(1 - \exp(-\lambda_{116m1} t_c)) \times B.R_{416.9\text{ keV}} \times \epsilon_{416.9\text{ keV}}} \qquad (7)$$

Taking the ratio of rates (Eq. 7) between the case with polyethylene plate in place to that without it, and noting that irradiation time is the same, gives:

$$\frac{R_{poly}}{R_{no\,poly}} = \frac{N_{poly} \times \exp(-\lambda t_{d,no\,poly})}{N_{no\,poly} \times \exp(-\lambda t_{d,poly})} \times \frac{(1 - \exp(\lambda t_{c,no\,poly}))}{(1 - \exp(\lambda t_{c,poly}))} \qquad (8)$$

Before the experimental values can be compared to the values from simulation, Eq. 8 needs to be normalized to the respective mass of the indium foil and the deuteron beam currents. The normalization factors are thus: $\frac{m_{foil,no\,poly}}{m_{foil,poly}}$ and $\frac{I_{poly}}{I_{no\,poly}}$. The mass factor ranges from 0.419 to 2.185. The average deuteron current for the first experiment in which there are polyethylene plates surrounding the neutron generator was 8.2 mA, while that for the second experiment in which the polyethylene plates were removed was 15.1 mA. Table 6 shows the experimental measurement of $\frac{R_{poly}}{R_{no\,poly}}$ after the normalization.

| Foil number | Polyethylene/no polyethylene | Absolute error |
| --- | --- | --- |
| | | |
| 1 | 6.887 | 0.696 |
| 2 | 4.305 | 0.417 |
| 3 | 5.704 | 0.843 |
| 4 | 6.467 | 0.821 |
| 5 | 3.198 | 0.438 |
| 6 | 3.927 | 0.508 |
| 7 | 6.442 | 0.825 |
| 8 | 4.734 | 0.686 |
| 9 | 5.172 | 0.874 |

Table 6. The ratio of $^{116m1}$In production rate for when polyethylene plate is present and when absent.

## V. RESULTS AND DISCUSSION

### V.A. Angle Differential Results

Taking the rate ratio of simulation to experiment (Table 5 and Table 6), Table 7 is obtained.

| Foil number | Simulation/Experiment | Absolute error |
|---|---|---|
| 1 | 0.93 | 0.11 |
| 2 | 1.20 | 0.13 |
| 3 | 1.04 | 0.16 |
| 4 | 0.74 | 0.10 |
| 5 | 1.07 | 0.15 |
| 6 | 1.19 | 0.16 |
| 7 | 1.01 | 0.14 |
| 8 | 1.01 | 0.16 |
| 9 | 1.12 | 0.20 |

Table 7. Simulation to experimental ratio of $^{116m1}$In rate of formation.

A plot of Table 7, showing simulation/experiment vs foil number is shown in Figure 7.

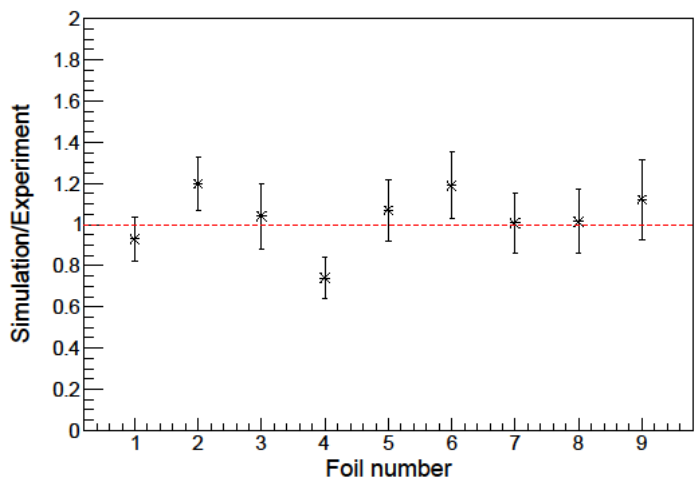

Figure 7. Comparison between simulation and experiment. The dashed red horizontal line represents the ideal case in which experiment and simulation agree completely. The error bar is that of statistical error.

From Figure 7, the simulation to experiment ratio agrees with the expected value of unity within the limit of statistical error, showing that the scattering data of hydrogen and carbon show fidelity to what is expected in a scattering experiment in the D-D neutron spectrum.

**V.B.  Angle integrated results**

The average angle integrated value of the ratio of simulation to experiment is:

$$\left(\frac{S}{E}\right)_{Avg} = 1.034 \pm 0.049 \tag{9}$$

The uncertainty estimate that needs to be added to the MCNP simulation result is then given by:

$$\left(\frac{E-S}{E}\right)_{Avg} = 1 - \left(\frac{S}{E}\right)_{Avg} = -0.034 \tag{10}$$

This is 3.4% and is within the uncertainty in the experimental measurements in Table 7. Thus, this again shows that the evaluated scattering cross section data of hydrogen and carbon in polyethylene reproduced very well what was observed in experiment for neutron energies in the spectrum of D-D neutron generator.

**V.C.  Systematic error**

The major source of systematic error was the alignment of the indium foils in respect to deuteron beam direction. In an ideal situation, the central indium is supposed to be centered on an axis defined by the deuteron beam direction. However, it is difficult to assure it. Hence, this lack of knowledge of the precise alignment of the indium foils introduces systematic error. Since neutron yield has a quadratic dependence on angle [8], the uncertainty in foil alignment can be estimated by a quadratic surface fit to the neutron flux at each foil location when there was a no polyethylene plate covering the HFNG. Using the uncertainty, the position of the indium foils in the MCNP simulation were adjusted and simulations carried out to determine the upper and lower limits of the systematic error. *Figure 8* shows the plot of the systematic error superimposed on the statistical error.

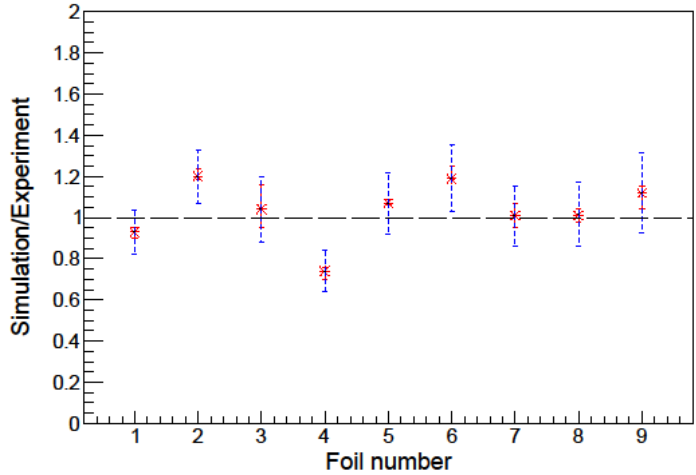

Figure 8. Comparison between experiment and simulation. The dashed horizontal line represents the ideal case in which experiment and simulation agree completely. The solid red error bars are the systematic errors while the dashed blue error bars are the statistical error

## VI    Conclusions

Using radiative capture reaction in indium foils, we have shown that the evaluated neutron scattering cross section data for the elemental components of polyethylene: hydrogen and carbon, agree with experimental measurement within limits of statistical error. The results from these foils are all within one sigma of the expected value of unity. Using multiple foils at different angles ensures that different neutron energies in the primary spectrum are considered and also decreased the possibility of compensating errors influencing the results.

The use of multiple foils presented a challenge: because only a single HPGe was used in collecting the spectra, the short half-life of $^{116m1}$In made it impractical to count the foils for a long time. A short period of 5 minutes had to be used. This resulted in statistical error in Table 4 being higher. In order to reduce this statistical error, each indium foil needs to be counted for a longer time. The best approach is simultaneous counting of all the foils. We suggest that a future work should take this into consideration.

Literature search shows that there has not been any work on benchmarking polyethylene using D-D generator neutron energy spectrum. Thus, our work appears to be the first to use compact D-D neutron generator to benchmark polyethylene. Polyethylene is an important material in the nuclear industry particularly in shielding and moderation of fast neutron to thermal neutron spectrum, via elastic scattering. By showing that the evaluated data on elastic scattering on its constituent elements accurately reproduces what is observed in experiment, nuclear data community are assured that their evaluation of neutron scattering cross sections in hydrogen and carbon are accurate within the limit of experimental error. In addition, users of polyethylene for shielding work are assured of high fidelity in the MCNP simulation.

This result also highlights the utility of compact D-D neutron sources such as the HFNG for integral benchmark experiments.

The authors are aware that ENDF/B-VIII.0 has been released. When the work was started, however, ENDF/B-VIII was in beta version and so was not used.

**Acknowledgement**

This work was performed under the auspices of the US DOE US Nuclear Data Program at LBNL under contract DE-AC02-05CH11231 and NSF grant EAR-0960138. The simulation used the Savio computational cluster resource provided by the Berkeley Research Computing program at the University of California, Berkeley (supported by the UC Berkeley Chancellor, Vice Chancellor for Research, and Chief Information Officer).